\documentclass[final,3p,times]{elsarticle}
\usepackage{amssymb, amsmath}
\usepackage{lipsum}

\journal{Annals of Physics}

\begin{document}

\begin{frontmatter}

\title{Classification of the  non-null electrovacuum solution of Einstein-Maxwell equations  with  three-parameter abelian group of motions}

\author[a,b]{V.V. Obukhov}
\affiliation[a]{organization={Tomsk State  Pedagogical University}, {Institute of Scietific Research and
Development,},
            addressline={60 Kievskaya St.},
            city={Tomsk},
            postcode={634041},
            state={Russia},
            country={Russian Federation}}
\affiliation[b]{organization={Tomsk State University of Control Systems and Radio Electronics}, {Laboratory for Theoretical Cosmology, International Center of Gravity and Cosmos,},
	addressline={36, Lenin Avenue},
	city={Tomsk},
	postcode={634050},
	state={Russia},
	country={Russian Federation}}

\begin{abstract}
The classification of the Stackel spaces of the electrovacuum of the type (3.0) has been done. These spaces are invariant under the action of the three-parameter abelian group of motions and belong to the first type Bianchi spaces.  In the case of a non-zero cosmological term, the metrics and potentials contain solutions of a nonlinear ordinary differential equation of the second order. When the cosmological term equals zero, the metrics and the components of the electromagnetic field tensor are expressed through elementary functions.Thus the classification of the electrovacuum stackel spaces of all types is completed and complete list of these spaces is constructed.
\end{abstract}


\end{frontmatter}

\section{Introduction}
\qquad
At present, a large number of exact solutions of Einstein equations are known.  Although some of them are important for the understanding of the nature of gravitation and for the development of mathematical methods of the General Relativity, the physical meaning of most of the known exact solutions is still not clear.  In  \cite{1}  authors have noted that it is not always easy to evaluate the usefulness of seemingly un physical specific exact solutions, since their possible qualitative properties are usually difficult to reveal because of the nonlinearity of the field equations of the General Relativity. At the same time, the noted nonlinearity in a number of cases significantly complicates the correct application of perturbation theory methods, so exact solutions can be very useful for checking the suitability of the approximate methods used.

Let's refine the term "exact solutions"
In the book \cite{1}, it has been proposed  to use this term in cases when the components of the metrics and electromagnetic potentials in some coordinate systems are given by means of known analytic functions, and also if these components are given in quadratures or with accuracy to solutions of some ordinary differential equation.

Although the problem of searching for new exact solutions of Einstein equations has largely lost its relevance, the classification problems is still of great importance, since classification allows one to identify and organize exact solutions of gravitational equations according to understandable physical and mathematical conditions. Four main methods of classification of exact solutions of Einstein's equations (including known ones) are noted:

\quad

- algebraic classification of the Weyl tensor (classification by Petrov [2]);

- algebraic classification of the Ricci tensor  and
by physical characteristics of the energy-momentum tensor;

- classification by structure of existing vector fields;

- classification by symmetry groups  and symmetry algebras of a given metrics and physical fields.

- embedding (see, for example, \cite{1a} \cite{1b}).
From the point of view of symmetry theory the last type of classification is of the greatest interest. In A.Z. Petrov's book \cite{2} the classification problem is completely solved for gravitational fields of general form admitting groups of motions. We will call such spaces Petrov spaces. In \cite{2} Petrov has also carried out a partial classification for Ricci flat spaces as well as for the corresponding Einstein spaces. A direct extension of this classification problem is the problem of classifying metrics of spacetime and of electromagnetic potentials in which the classical and quantum equations of motion of charged test particles admit algebras of integrals of motion linear or square in momentum. In these cases spaces admit complete sets of mutually commuting vector and tensor Killing fields.
Such spaces, are called Stackel spaces. In Stackel spaces, the geodesic equations can be integrated by the method of complete separation of variables in the free Hamilton-Jacobi equation. The Lorentz, Klein-Gordon-Fock, Dirac-Fock and Weyl equations can be integrated by the method of complete separation of variables for some subsets of the set of stackel spaces. In this case, the quantum mechanical and wave equations are reduced to systems of ordinary differential equations. The finding of these subsets is the aim of the corresponding classifications. It should be noted that the eikonal equation can be integrated by the method of complete separation of variables on a conformal-Stackel manifold (the wave equation - on some subsets of this manifold).

Let us clarify what is meant by the term "classification" in this case.
The main theorems of the theory of separation of variables have been proved by V.N. Shapovalov (\cite{3},\cite{4},\cite{5}). It  have been established  the covariant features of the separation of variables in the equations of motion of a test particle. It has been proved that a complete separation of variables is possible if and only if the complete sets of integrals of motion (symmetry operators) forming a commutative algebra exist. Thus the existence of a complete set consisting of appropriate vector and (or) tensor Killing fields  is a necessary condition for complete separation of variables. Spaces that have complete sets have been called Stackel spaces. In Stackel spaces there exist "privileged" coordinate systems.  Complete separation of variables is possible only in these systems.

For the charged Hamilton-Jacobi equation, the Klein-Gordon-Fock, Dirac-Fock, and Weyl equations, there are additional restrictions both on the metric of the stackel space and on the electromagnetic potentials (see \cite{5}, \cite{6}, \cite{7}, \cite{8}). In privileged coordinate systems, Killing vector and tensor fields are "diagonalized". There exist groups of admissible transformations of privileged coordinate systems that do not break the diagonalization. Metrics linked by admissible transformations form equivalence classes. The purpose of the classification is to list all these classes. In the presence of an electromagnetic field, the group of admissible transformations is supplemented by a gradient transformation of the electromagnetic potential. The first works on the classification of Stackel metrics (\cite{9} -\cite{12}) were devoted to the metrics of flat space given in curved coordinates. The result of such classification was the complete list of all privileged coordinate sets and complete sets of integrals of motion of the charged  Klein-Gordon and Dirac equations.

The solution of classification problems for stackel spaces and Petrov spaces is not only of purely mathematical interest, but also of physical interest, since the possibility of obtaining exact solutions of the motion equations for test particles gives additional methods both for studying the geometry of the corresponding spaces and for researching the physical processes occurring in them. The aim of classification is to construct complete lists of such fields. The importance of carrying out such an activity is evidenced by the following circumstances.

1) Almost all the most interesting from the physical point of view exact solutions of Einstein equations belong to the number of Stackel or Petrov spaces.

 Note that the interest in these spaces has not waned until now. As an example, we can point to the papers in the field of cosmology, including the modified theory of gravitation \cite{19}, \cite{20}, These are such solutions as the static spherically symmetric Schwarzschild solution \cite{13},\cite{14}, the solution for a rotating Kerr source \cite{15}, for a static charged sphere Reissner-Nordstrem \cite{16}, for a rotating Newman-Unti-Tamburino charge \cite{17}, cosmological Friedmann solutions \cite{18} and some others. \cite{21}, \cite{22}, \cite{23}, \cite{25}. A separate direction of application of stackel spaces is their use in multidimensional gravitation (see, \cite{26}, \cite{27}, \cite{28}).

2) Identification of the Stackel spaces from the set of already known exact solutions of the Einstein equations is complicated. These solutions can be represented in non-privileged coordinate systems, and their identification as stackel solutions is usually a non-trivial problem. Moreover, until a complete classification of the corresponding solutions of the field equations has been carried out, it is impossible to claim that all  exact solutions, in which the method of complete separation of variables is applicable, are known. The same remarks are also true for the Petrov spaces.

Interest in Petrov spaces increased after the advent of the method of non-commutative integration. The method can be used for the Klein-Gordon-Fock and Dirac-Fock equations in space-time manifolds with three or four parametric groups of motions acting simply transitively on the manifold itself or on its three-dimensional submanifold (see, for example, \ cite{29}). The non-commutative integration method leads to the reduction of motion equations to systems of ordinary differential equations. This circumstance unites Petrov spaces and Stackel spaces into a common set spaces, in which it is possible to reduce the quantum mechanical motion equations to systems of ordinary differential equations. In order to emphasize this common feature, the method of complete separation of variables is also called the method of commutative integration. If, in addition to the complete set of Killing fields, a stackel space has additional vector Killing fields, then it belongs to the set which is the intersection of the set of Stackel spaces and the set of Petrov spaces. These intersections were studied in the papers \cite{30}, \cite{31}, \cite{32}.

In the papers \cite{33}, \cite{34}, \cite{35}, \cite{36}, \cite{37}, \cite{38} (see also the presented bibliography) the classification of electrovacuum Stackel spaces was carried out (including with a cosmological term) for all types of Stackel metrics with the exception of the case, when space and the electromagnetic field admit a non isotropic complete set consisting of three mutually commuting Killing fields. According to the terminology adopted in the theory of complete separation of variables, such spaces are called Stackel spaces of type (3.0).or (3.1) \cite{32}.

The first stage of the problem of classifying Petrov spaces was solved by Petrov in the book \cite{2}. All non-equivalent space-time metrics in which the motion equation for a classical uncharged particle admits the corresponding groups of linear on momentum integrals of motion has been listed.

At the second stage of solving the classification problem all non-equivalent potentials of admissible electromagnetic fields are found.
In the papers \cite{39}, \cite{40}, \cite{41}, \cite{42} this problem has been solved for the case, when the linear on momentum motion integrals of charged Hamilton-Jacobi and Klein-Gordon-Fock equations form a four-parametric group. It has been established that this group coincides with the simply transitively four-parametric group of motions acting in space. In the papers \cite{43}, \cite{44}, \cite{45} all admissible electromagnetic fields for Petrov spaces with simply transitive three-parametric groups of motions were classified.

At the third stage, it is necessary to classify the exact solutions of the Maxwell and Einstein-Maxwell equations for Petrov spaces. For homogeneous Petrov spaces with a four-parameter group of motions, the problem was solved in the papers \cite{46}, \cite{47}, \cite{48}, \cite{49}. In the papers \cite{50}, using numerical methods, solutions of the Einstein-Maxwell equations for Petrov spaces with the Bianchi type IX group of motions are obtained. For the remaining Petrov spaces, exact solutions of source-free Maxwell equations have been classified in the papers \cite{51}, \cite{52}, \cite{53}, \cite{54}, \cite{55}.

The present paper is devoted to the complete classification of non-isotropic Stackel spaces of the electrovacuum of type (3.0). These spaces also belong to Petrov spaces of the first type according to Bianchi classification. Thus, while completing the classification for the set of electrovacuum Stackel spaces, we simultaneously begin the classification of electrovacuum Petrov spaces. Note that all Ricci flat spaces as well as Einstein spaces with Stackel metrics of type (3.0) are already known either as special cases of electro-vacuum stackel spaces of type $(N. N_0)$ (at $N<3$) or obtained directly from Einstein's equations in \cite{56}, \cite{57}, \cite{58}, \cite{59}. As for the electrovacuum Stackel spaces of type (3.0), only one of them has been explicitly identified so far (see \cite{60}).

\section{Einstein-Maxwell vacuum equations}
Let us consider the vacuum Einstein-Maxwell equations.
\begin{equation} \label{1_1}
R_{ij} -\frac{1}{2} g_{ij} R=8\pi\kappa T_{ij} +g_{ij} \lambda,
\end{equation}
\begin{equation} \label{1_2_}
\frac{1}{\sqrt{\left|g\right|} }(F^{ij} \sqrt{\left|g\right|})_{,j} =0.
\end{equation}
Here $R_{ij} $ is the Ricci tensor:
\begin{equation} \label{1_3_}
R_{ik} =\Gamma _{ik,\ell }^{\ell } -\Gamma_{\ell i,k}^{\ell } + \Gamma_{ik}^{\ell} \Gamma _{\ell j}^{j} -\Gamma_{ij}^{\ell } \Gamma_{k\ell }^{j}, \quad R=g^{ik} R_{ik;}
\end{equation}
${T} _{ij} $ -- the energy-momentum tensor of the electromagnetic field:
\begin{equation} \label{1_4_}
{T} _{ij} =\frac{1}{4\pi }(-F_{i.}^{. \ell} F_{\ell j} +\frac{1}{4} g_{ij} F_{k\ell } F^{k\ell }),
\end{equation}
$F_{ij} $ is the electromagnetic field tensor:
\begin{equation} \label{1_5_}
F_{ij} ={\rm A}_{j,i} -{\rm A}_{i,j,}
\end{equation}
$g_{ij} $ is the metric space-time tensor $V_{4} $,\quad $\lambda $ is the cosmological term.

Let the metric and the electromagnetic field be invariant with respect to the abelian group of motions given by the triple of Killing vectors:
\begin{equation} \label{1_6_}
\xi _{\alpha }^{i} =\delta _{\alpha }^{i}.
\end{equation}
Then there exists a privileged coordinate system $\left\{u^{i} \right\}$, in which the components of the metric tensor $g_{ij}$ and the electromagnetic potential $A_{i} $ depend only on the variable $u^{0}$. We will consider the case, when the group acts simply transitively on a non-null hypersurface of the space $V_{4}$. In this case a semi-geodesic coordinate system can be chosen as the privileged coordinate system, and  metric tensor $g_{ij}$  will take the form:
\begin{equation} \label{1_7_}
g_{ij} =-\varepsilon \delta _{i}^{0} \delta _{j}^{0} +\delta _{i}^{\alpha } \delta _{j}^{\beta } \eta _{\alpha \beta } \left(u^{0} \right)du^{\alpha } du^{\beta } .\qquad \qquad \left(\varepsilon =\pm 1\right)
\end{equation}
If $\varepsilon =1$, the variable $u^{0}$ is temporal. If $\varepsilon =-1,\quad u^{0}$ is a space variable.
Applying the gradient transformation, we reduce the electromagnetic potential to the form
\begin{equation} \label{1_8_}
{\rm A} _{i} =\delta _{i}^{\alpha } { A} _{\alpha } \left(u^{0} \right).
\end{equation}
Next notations are used here and hereafter:
\begin{equation} \label{1_9_}
i,j,k,\ell=0,1,2,3;\qquad \alpha ,\beta ,\gamma =1,2,3;\qquad p,q,r=2,3.
\end{equation}
The repeating upper and lower indices are summed up within the established for them limits for indices changes. The determinant of the matrix $|| \eta _{\alpha \beta } || $ can be represented as:
\begin{equation} \label{1_20_}
\det{|| \eta _{\alpha \beta } ||} =\varepsilon \ell ^{2} \qquad \left(\ell =\ell \left(u^{0} \right)\right).
\end{equation}
Note that the Hamilton-Jacobi equation
\begin{equation}\label{1c}
g^{ij}(S_{,i}+{\rm A}_i)(S_{,j}+{\rm A}_j)=m^2,
\end{equation}
and Klein-Gordon-Fock equation
\begin{equation}\label{2c}
g^{ij}(-\imath\nabla{,i}+{\rm A}_i)(-\imath \nabla_{,j}+{\rm A}_j)\psi=m^2\psi
\end{equation}
can indeed be integrated by the method of complete separation of variables. Let us show this on the example of the Hamilton-Jacobi equation \eqref{1c}. If the solution of the Hamilton-Jacobi equation can be represented as
\begin{equation}\label{3c}
S = f(u^0, \lambda_i) + \lambda_\alpha u^\alpha
\end{equation}
($\lambda_i $ are essential parameters, $\lambda_0 = m^2$),
then the function \eqref{3c} is the complete integral of the Hamilton-Jacobi equation \eqref{1c}, and the Lorentz equations can be integrated by the Hamilton-Jacobi method. Let us substitute \eqref{3c} into the \eqref{1c} equation. We obtain an ordinary differential equation on the function $f$:
\begin{equation}\label{4c}
\varepsilon (f_{,0})^2 -\eta^{\alpha\beta}A_\alpha A_\beta + m^2 = 0.
\end{equation}
 Let us substitute the functions ${\rm A} _{i} $ from the expression \eqref{1_8_} into \eqref{1_5_}. The result is:
\begin{equation} \label{1_21_}
F_{0\alpha } ={A} _{\alpha ,0},\quad F_{\alpha \beta } =0\Rightarrow F^{0\alpha } =-\varepsilon \eta ^{\alpha \beta } {A} _{\beta ,0} =-\varepsilon \beta ^{\alpha },
\end{equation}
where it is denoted \quad $\beta ^{\alpha } =\eta ^{\alpha \beta } {A} _{\beta ,0}$. Maxwell's equations can be written as:

\begin{equation} \label{1_22_}
(\ell \beta^{\alpha })_{,0} =0\Rightarrow \beta ^{\alpha } =\frac{P^{\alpha } }{\ell } \qquad \left(P^{\alpha } =const\right)
\end{equation}
The group of admissible transformations of privileged coordinate systems is given by the matrix $\hat{S}:$
\begin{equation} \label{1_23}
\widetilde{u}^{\alpha } =S_{\beta }^{\alpha } u^{\beta },
\end{equation}
where $S_{\beta }^{\alpha } $  is an arbitrary non singular matrix with constant elements:
\begin{equation} \label{1_24_}
S_{\alpha }^{\beta} =const,\quad \det|S_{\alpha }^{\beta}| \ne 0.
\end{equation}
Using the \eqref{1_23} transformation, the solution of Maxwell equations \eqref{1_22_} can be represented as:
\begin{equation} \label{1_25_}
\beta ^{\alpha } =\delta _{1}^{\alpha } \left(\frac{\alpha }{\ell } \right),\qquad \alpha =const
\end{equation}
Indeed, let us represent the parameters $P^{\alpha } $ in the form:
$$ \label{1_26_}
P^{\alpha } =B_{1}^{\alpha } \alpha .
$$
If we choose constants $B_{1}^{\alpha } $ as elements of the first row of the matrix
$\left(\widehat{S}^{-1} \right)_{\beta }^{\alpha } $, from \eqref{1_23} follows:
\[\widetilde{P}^{\alpha } =S_{\beta }^{\alpha } P^{\beta } =S_{\beta }^{\alpha } (S^{-1}) _{1}^{\beta } \alpha =\delta _{1}^{\alpha } \alpha .\]
The subgroup of the of admissible transformations group that does not break the condition \eqref{1_25_} is given by the matrix:
\begin{equation}\label{1_26_}
\hat{S}=\begin{pmatrix}1 & 0 & 0 \\ S_2^1 & S_2^2 & S_2^3 \\  S_3^1 & S_3^2 & S_3^3  \end{pmatrix}
\end{equation}
Let us substitute \eqref{1_25_} into the mixed components $T_{j}^{i} $ of the energy-momentum tensor \eqref{1_4_}. As a result, we obtain:
\begin{equation} \label{1_27_}
T_{0}^{0} =\frac{\varepsilon \kappa \alpha ^{2} }{4\pi \ell ^{2} } \eta _{11} ,\qquad T_{\beta }^{\alpha } =\frac{\varepsilon \kappa \alpha ^{2} }{4\pi \ell ^{2} } \left(2\delta _{1}^{\alpha } \delta _{\beta }^{1} -\delta _{\beta }^{\alpha } \right)
\end{equation}
The mixed components $R_{j}^{i} $ of the Ricci tensor are given in the book \cite{61}
\begin{equation} \label{1_28_}
R_{0}^{0} =\frac{\varepsilon }{4} \left(2\widetilde{\kappa }_{\alpha ,0}^{\alpha } +\widetilde{\kappa }_{\beta }^{\alpha } \widetilde{\kappa }_{\alpha }^{\beta } \right),\qquad R_{\beta }^{\alpha } =\frac{\varepsilon }{2\ell } \left(\ell \widetilde{\kappa }_{\alpha }^{\beta } \right)_{,0},
\end{equation}
where $\widetilde{\kappa }_{\beta }^{\alpha } =\eta ^{\alpha \gamma } \eta _{\gamma \beta ,0}. $ Using the relations \eqref{1_27_}, \eqref{1_28_}, the equations \eqref{1_1} can be represented as:
\begin{equation}\label{1_30_}
\left\{\begin{array}{c} {2\tilde{\kappa}_{\alpha,0}^{\alpha} +\tilde{\kappa }_{\beta }^{\alpha}  \tilde{\kappa}_{\alpha }^{\beta } =\frac{\kappa\alpha^2}{\pi \ell^2}\eta_{11} -8\varepsilon \lambda,} {}
\\ {\frac{1}{\ell } (\ell \tilde{\kappa}_{\beta }^{\alpha})_{,0} =\frac{\varepsilon \kappa \alpha ^{2} }{2\pi \ell ^{2} } \eta _{11}(2\delta _{1}^{\alpha } \delta _{\beta }^{1} -\delta _{\beta }^{\alpha }) -4\varepsilon \lambda \delta _{\beta }^{\alpha }.} \end{array}\right. \
\end{equation}
Let us transform the system of equations \eqref{1_30_} as follows.

1.
We introduce a new variable $u^{0} =u^{0} (\tau ).$ The function $u^{0} $ obeys the condition:
$$
\frac{du^{0}}{d\tau } =\ell(\tau).
$$
The derivatives on the variable $\tau $ will be denoted by points.

2.
We denote:
\begin{equation}\label{1_31_}
\kappa _{\beta }^{\alpha } =\eta^{\alpha \gamma } \dot{\eta }_{\gamma \beta }.
\end{equation}

3.
 We introduce new functions $\beta(\tau),\gamma(\tau)$ satisfying the equations:
\begin{equation} \label{1_32_}
\left\{\begin{array}{c} {\ddot{\beta }=\frac{\kappa \alpha ^{2} }{2\pi }\eta _{11}.}
\\{}
\\ {\ddot{\gamma }=-4\varepsilon \lambda \ell ^{2} = 4\xi|\lambda|\ell ^{2},}\end{array}\right.
\end{equation}
Obviously, for electro-vacuum case  $\ddot{\beta } \ne 0$, otherwise we obtain a vacuum Stackel space-time.
Einstein-Maxwell equations \eqref{1_30_} can be represented as:
\begin{equation} \label{1_33_}
2\dot{\kappa }_{\alpha }^{\alpha } +\kappa _{\beta }^{\alpha } \kappa _{\alpha }^{\beta } -\left(\kappa _{\alpha }^{\alpha } \right)^{2} =2\left(\ddot{\beta }+\ddot{\gamma }\right).
\end{equation}
\begin{equation} \label{1_34_}
\dot{\kappa }_{\beta }^{\alpha }=2\ddot{\beta }(\delta_{1}^{\alpha} \delta_{\beta}^{1})+\delta_{\beta }^{\alpha } (-\ddot{\beta}+\ddot{\gamma}).
\end{equation}
From equations \eqref{1_34_} it follows:
\begin{equation} \label{1_35_}
\kappa _{\beta }^{\alpha } =2\dot{\beta }(\delta _{1}^{\alpha } \delta _{\beta }^{1}) +\delta _{\beta }^{\alpha } \left(\dot{\gamma }-\dot{\beta }\right)+C_{\beta }^{\alpha } \qquad \left(C_{\beta }^{\alpha } =const\right).
\end{equation}
Using \eqref{1_35_} \eqref{1_31_}, we obtain the system of equations:
$$
\dot{\eta }_{\alpha \beta } =2\eta _{\alpha 1} \delta _{\beta }^{1} \dot{\beta }+\eta _{\alpha \beta } \left(\dot{\gamma }-\dot{\beta }\right)+C_{\beta }^{\alpha } \eta _{\alpha \gamma }.
$$
Let us denote:
\begin{equation} \label{1_36_}
\eta _{\alpha \beta } =\omega _{\alpha \beta } \exp \left(\gamma -\beta \right).
\end{equation}
Then the system \eqref{1_35_} can be represented as: 
\begin{equation} \label{1_37_}
\left\{\begin{array}{c} {\dot{\omega }_{11} =2\dot{\beta }\omega _{11} +C_{1}^{\alpha } \omega _{1\alpha },}\qquad \qquad \qquad \qquad \qquad \qquad \qquad \qquad \qquad \\

 {\dot{\omega }_{1p} =2{\dot{\beta }} \omega _{1p} +C_{1}^{\alpha } \omega _{p\alpha }.}\qquad \qquad \qquad \qquad \qquad \qquad \qquad \qquad \qquad \\

{\dot{\omega }_{p1} =C_{p}^{\alpha } \omega _{\alpha 1} \qquad \qquad \qquad \qquad \qquad \qquad \qquad \qquad \qquad \qquad \qquad  } \\

 {\dot{\omega }_{pp} =C_{p}^{1} \omega _{1p} +C_{p}^{q} \omega _{pq} \qquad \qquad \qquad \qquad \qquad \qquad \qquad}\qquad \qquad \\

 {\dot{\omega }_{pq} =C_{q}^{1} \omega_{1} +C_{q}^{r } \omega _{pr} \quad p \ne q. \qquad \qquad \qquad \qquad \qquad  \qquad \qquad } \end{array}\right.  \
\end{equation}
Let's introduce the notations:
\begin{equation} \label{1_42_}
C_{2}^{p} =a^{p},\quad C_{3}^{p} =b^{p} ,\quad C_{1}^{\alpha} =c^{\alpha}, \quad
\end{equation}
$$ C_{p}^{1} =c_{p},\quad\sigma =2\dot{\beta }+c^{1}$$.
\begin{equation} \label{1_43_}
\mathbb{A}_{p} =a^{q} \omega _{qp} ,\qquad \mathbb{B}_{p} =b^{q} \omega _{qp} ,\qquad \mathbb{P}_{p} =c^{q} \omega _{qp}
\end{equation}
Instead of \eqref{1_37_}, we get (using: $\dot{\omega}_{\alpha\beta} = \dot{\omega}_{\beta\alpha}$) the system of equations:
\begin{equation}\label{1_44}
\left\{\begin{array}{c} {\dot{\omega }_{11} =\sigma \omega _{11} +c^{p} \omega _{1p} \qquad \qquad \qquad  } \\ {\dot{\omega }_{1p} =\sigma \omega _{1p} +\mathbb{P}_{p} \qquad \qquad \qquad }
\\ {\dot{\omega }_{22} =c_2 \omega _{12} +\mathbb{A}_{2} \qquad \qquad \qquad  }
\\ {\dot{\omega }_{33} =c_3\omega _{13} +\mathbb{B}_{3} \qquad \qquad \qquad }
\\ {\dot{\omega }_{23} =c_3 \omega _{12} +\mathbb{B}_{2} \qquad \qquad \qquad  } \end{array}\right. \
\end{equation}
\begin{equation}\label{1_45}
\left\{\begin{array}{c} {c_{3} \omega _{12} -c_{2} \omega _{13} =\mathbb{A}_{3}  - \mathbb{B}_{2} \qquad \qquad \qquad }
\\ {\left(\sigma -a^{2} \right)\omega _{12} -a^{3} \omega _{13} =c_{2} \omega _{11} -\mathbb{ P}_{2} \qquad \qquad } \\ {-b^{2} \omega _{12} +\left(\sigma -b^{3} \right)\omega _{13} =c_{3} \omega _{11} - \mathbb{P}_{3} \qquad \qquad } \end{array}\right. \
\end{equation}
The systems of equations \eqref{1_44}, \eqref{1_45} together with the system \eqref{1_32_} form the complete system of the Einstein-Maxwell electro-vacuum equations.
The system of equations \eqref{1_44} -- \eqref{1_45} does not impose restrictions on the function $\gamma $. In the next section we will show that it does not impose restrictions on the $\beta $ function either. Therefore, we will call it an autonomous system. The autonomous system includes a subsystem of linear algebraic equations \eqref{1_45} linking the functions $\omega_{\alpha\beta}$. Let us find the determinant of the matrix
 \begin{equation} \label{3_1}
\Delta = \det \begin{pmatrix} c_3 & -c_2 & 0 \\
 \sigma-a^2 & -a^3 & -c_2 \\
     -b^2 & \sigma-b^3 & -c_3
\end{pmatrix} =
\end{equation}
$$
= \left(c_{3} \right)^{2} a^{3} -\left(c_{2} \right)^{2} b^{2} +c_{2} c_{3} \left(a^{2} -b^{3} \right).
$$
We will consider the solution of the autonomous system separately for the variants $\Delta \ne 0$, and $\Delta =0$. In both cases, as shown below, $\omega_{1p}=0.$

\section{ Classification of the matrix $\omega_{\alpha\beta} $}

\subsection{Solution of the autonomous system at $\Delta \ne 0$}
When $\Delta \ne 0$ to find the functions $\omega_{1p}$ we have an algebraic system of equations, which is a consequence of the system of equations \eqref{1_45}:
\begin{equation} \label{3_2}
\left\{\begin{array}{c} {c_{3} \omega _{12} -c_{2} \omega _{13} =\mathbb{A}_{3} -\mathbb{B}_{2} } \\ {\left[c_{3} \left(\sigma -a^{2} \right)+c_{2} b^{2} \right]\omega _{12} -\left[c_{2} \left(\sigma -b^{3} \right)+c_{3} a^{3} \right]\omega _{13} =
} \end{array}\right.
\end{equation}
$$=c_{2} \mathbb{P}_{3} -c_{3} \mathbb{P}_{2} $$
Since the determinant of this system does not equal to zero, the solution has the form:
\begin{equation} \label{3_3}
\left\{\begin{array}{c} {\omega _{12} =\frac{1}{\Delta}(c_2((\sigma-a^2 -b^3)(\mathbb{A}_{3} -\mathbb{B}_{2,})}+ \\ {+c_3\mathbb{P}_2 -c_2\mathbb{P}_3)+(\mathbb{A}_{3} -\mathbb{B}_{2,})c_pa^p),} \\{}
\\ {\omega _{13} =\frac{1}{\Delta}(c_3((\sigma-a^2 -b^3)(\mathbb{A}_{3} -\mathbb{B}_{2,}) +} \\{c_3\mathbb{P}_2 -c_2\mathbb{P}_3)+(\mathbb{A}_{3} -\mathbb{B}_{2,})c_pb^p)}. \end{array}\right.
\end{equation}
The system of equations (37) is linear in functions $\omega_{ab}$.
 It can be represented as:
$\omega_{1p} = W_p^{q\tilde{p}} \omega_{q \tilde{p}}$ The elements of the matrix $W_p^{q\tilde{p}}$ are composed from the coefficients in front of the functions $\omega _{pq}$. At admissible coordinate transformations of the form (20), the functions $\omega _{\alpha\beta}$ are transformed as follows
$$
\tilde{\omega}_{\alpha\beta}= S_\alpha^{\tilde{\alpha}}S_\beta^{\tilde{\beta}}{\omega_{\tilde{\alpha}\tilde{\beta}}}
$$
Let us choose these transformations in the form:
$$
\tilde{u}^1 = u^1, \quad \tilde{u}^{2} =(c_{3} a^{3} -c_{2} b^{3})u^{2} +(c_{2} b^{2} -c_{3} a^{2})u^{3}, \quad \tilde{u}^{3} =c_{2} u^{2} +c_{3} u^{3}.
$$
It is possible since $\Delta \ne 0.$ Then the solution of the system of equations \eqref{3_3} can be represented in the form:
\begin{equation} \label{3_4}
\left\{\begin{array}{c} {\omega _{12} =\mathbb{A}_{3} -\mathbb{B}_{2,} } \\ {\omega _{13} =\left(\mathbb{A}_{3} -\mathbb{B}_{2} \right)\sigma +\mathbb{P}_{2} }. \end{array}\right.
\end{equation}
Hence
$$
c_{3} =a^3 =1, \quad c_{2} = 0. 
$$
Let us differentiate the first equation of the system \eqref{3_4} by the variable $\tau $. Using the equations from the system \eqref{1_44} we obtain:
\[\dot{\omega }_{12} =\dot{\mathbb{A}}_{3} -\dot{\mathbb{B}}_{3} = \dot{\omega }_{33} +a^2\dot{\omega }_{23} -b^{2} \dot{\omega }_{22} -b^{3} \dot{\omega }_{23} =
$$
$$=\omega _{13} +\mathbb{B}_{3} -b^{2} \omega _{22} -b^{3} \left(\omega _{33} \right)=\sigma\omega _{12} +\mathbb{P}_{2}. \]
Hence:
\begin{equation} \label{3_5}
a^2 (\mathbb{A}_{3} -\mathbb{B}_{2} )=0.
\end{equation}
The following options are possible: $\mathbb{A}_{3}-\mathbb{B}_{2}=0$, or $a^2$. In the first variant $\omega_{12}=0$ (it follows from the first equation of the system \eqref{3_4}). From the second equation of the system \eqref{1_44} at $p=2$ we obtain $$\mathbb{P}_2 = 0 \rightarrow \omega_{13}=0.$$ Consider the variant $a^2=0$. Differentiating the second equation of the system \eqref{3_4} by $\tau$ and, comparing the resulting expression with the second equation of the system \eqref{1_44} at $p=3$, we obtain:
$$
\dot{\sigma}\omega_{12}=0.
$$
Since for electrovacuum solutions $\dot{\sigma} \ne 0$, from this equation it follows $\omega_{12}=0 \rightarrow \mathbb{A}_{3} -\mathbb{B}_{2}=0.$ We obtain the solution belonging to the first option. Thus, the system of equations \eqref{3_2} at $\Delta \ne 0$ has a solution:
$$
\omega_{12}=\omega_{13}=0.
$$
From the second equation of the system \eqref{1_44} we have: $P_q=0$. Then from the third equation of the system \eqref{1_45} it follows: \quad $\omega_{11} = 0,$ \quad which is impossible.
\subsection{Solution of the autonomous system at $\Delta = 0$}
$\bf{1.} \quad $ Consider the case where Let $ (c_{2})^{2} +(c_{3})^{2} \ne 0$. Without restriction of generality, we believe  $c_{3} \ne 0$. From the first two equations of the system \eqref{1_45} it follows:
\begin{equation} \label{3_6}
\omega _{12} =\frac{1}{c_{3} } \left(c_{2} \omega _{13} +\mathbb{A}_{3} -\mathbb{B}_{2} \right)
\end{equation}
\begin{equation} \label{3_7}
\left(\mathbb{A}_{3} -\mathbb{B}_{2} \right)\left(\sigma c_{3} +b^{2} c_{2} - a^{2}c_{3} \right)=c_{3} \left(c_{2} \mathbb{P}_{3} -c_{3} \mathbb{P}_{2} \right)
\end{equation}
From the last equation of the system \eqref{1_45} we obtain:
\begin{equation} \label{3 8}
\omega _{11} =\frac{1}{c_{3} } \left[\left(\sigma -b^{3} \right)\omega _{13} -b^{2} \omega _{12} +\mathbb{P}_{3} \right]
\end{equation}
The first equation of the system \eqref{1_44} can be written as:
\begin{equation} \label{3 9}
\dot{\omega }_{11} =\frac{\sigma}{c_{3} } \left[\left(\sigma -b^{3} \right)\omega _{13} -b^{2} \omega _{12} +P_{2} \right]+c^{2} \omega _{12} +c^{3} \omega _{13}
\end{equation}
Let us differentiate the equation \eqref{3 8} by $\tau$, and equate the resulting expression to the right-hand side of the equation \eqref{3 9}. Using the equations of the system \eqref{1_44}, we obtain:
$$
(\dot{\sigma}\omega_{13} +(\sigma -b^{3})(\sigma \omega_{13} +\mathbb{P}_{3}))
-b^{2}(\sigma (c_{2} \omega_{12} +\mathbb{A}_{3} -\mathbb{B}_{2} )+\mathbb{P}_{2} c_{3} )
$$
$$
+c_3((c^{2} c_{2} +c^{3} c_{3} )\omega _{13} +c^{2} \mathbb{A}_{3} +c^{3} \mathbb{B}_{3})=
\frac{\sigma }{c_{3}} (c_{3} (\sigma -b^{3} )\omega _{13}
-b^{2} (c_{2} \omega_{13} +\mathbb{A}_{3} -\mathbb{B}_{2})+c_{3} \mathbb{P}_{3} )+c_{3}^2 (c^{2} \omega_{12} +c^{3} \omega_{13}).
$$
Hence.
$$\dot{\sigma}\omega _{13} =0\Rightarrow \omega _{13} =0.$$
If $c_{2} \ne 0$, repeating the calculations (by substituting $2\leftrightarrow 3$), we obtain:
\begin{equation} \label{3_10}
\dot{\sigma }\omega _{12} =0 \rightarrow \omega _{12} =0.
\end{equation}
Therefore, for a final decision, we need to consider the case
$$
\omega _{13} =c_{2} =0.
$$
From the second equation of the system \eqref{1_44} at $p=3$ we obtain: \quad $\mathbb{ P}_{3} =0.$ \quad Since $\Delta =0, \quad c_3 \ne 0, \rightarrow a^{3} =0$.\quad We can believe that \quad $c_{3} =1$. Integrating the last three equations from the system \eqref{1_44}, we obtain the functions $\omega_{2p}$ ($\tilde{p}, \tilde{q} =const$):
\begin{equation}\label{3 11}
\omega _{22} =\tilde{p}\exp (a^{2} \tau ), \quad \omega _{23} =\tilde{q}\exp (a^{2} \tau ),
\end{equation}
As $\dot{\sigma} \ne 0$, the equation \eqref{3_7} is satisfied only if $\mathbb{A}_{3} -\mathbb{B}_{2}= \mathbb{P}_{2} =0$. Hence (from the equation \eqref{3_4}) it follows that $\omega _{12} =0$. From the last equation of the system \eqref{1_45} follows $\omega_{11}=0$, which is impossible. Thus, the system of equations \eqref{1_37_} must be considered under the conditions:
\begin{equation}\label{3 12}
\omega_{12}=\omega_{13}=c_p =c^q= 0.
\end{equation}
The last equality follows from the conditions:
\begin{equation}\label{3 13}
\mathbb{A}_{3} -\mathbb{B}_{2}=0, \quad \mathbb{P}_p =0.
\end{equation}

\quad
$\bf{2.}\quad $ Consider the case when $ c_{p}=0$. System \eqref{1_45} has the form:
\begin{equation}\label{1_13a}
\left\{\begin{array}{c} {\mathbb{A}_{3}  - \mathbb{B}_{2} =0 \qquad \qquad \qquad }
\\ {\left( a^{2} -\sigma\right)\omega _{12} + a^{3} \omega _{13} = \mathbb{ P}_{2} \qquad \qquad } \\ {b^{2} \omega _{12} +\left( b^{3}-\sigma \right)\omega _{13} = \mathbb{P}_{3} \qquad \qquad } \end{array}\right. \
\end{equation}
Hence
\begin{equation}\label{1_13b}
\left\{\begin{array}{c} {\omega _{12}=-\frac{1}{\Delta_1} \left(\Omega\mathbb{P}_2 +a^q\mathbb{P}_q\right),}
\\ {\omega _{13}=-\frac{1}{\Delta_1} \left(\Omega\mathbb{P}_3 +b^q\mathbb{P}_q\right);} \end{array}\right. \
\end{equation}
where \quad $\Delta_1 = (\Omega +a^2)(\Omega +b^3) -a^3b^2, \quad \Omega = \sigma -a^2 -b^3.$

Let us differentiate the equations \eqref{1_13b} by $\tau$, and equate the resulting expressions to the right-hand side of the equations \eqref{1_44}. Using the remaining equations of the system \eqref{1_44}, we obtain:
\begin{equation}\label{1 13c}
\left\{\begin{array}{c} {\dot{\Omega}((\Omega^2 + a^3b^2 -a^2b^3)\mathbb{P}_2 +(2\Omega +a^2 +a^3)\mathbb{P}_3) =0,}
\\ {\dot{\Omega}((\Omega^2 + a^3b^2 -a^2b^3)\mathbb{P}_3 +(2\Omega +a^2 +a^3)\mathbb{P}_2) =0,} \end{array}\right. \
\end{equation}
As \quad $\dot{\Omega} = \dot{\sigma} \ne const$, \quad from the system \eqref{1 13c} it follows:
$$\mathbb{P}_q=0 \rightarrow c^q =0 \rightarrow \omega_{1p}=0.$$
The autonomous system takes the form:
\begin{equation}\label{3 14}
 \dot{\omega }_{22} =\mathbb{A}_{2},
\quad \dot{\omega }_{23} =\mathbb{B}_{2},\quad \dot{\omega }_{33} =\mathbb{B}_{3},\quad
\dot{\omega }_{11} =\sigma \omega _{11},
\end{equation}
The first equation \eqref{3 13} can be represented as follows:
\begin{equation} \label{3_17}
\left(a^{2} -b^{3} \right)\omega _{23} +a^{3} \omega _{33} -b^{2} \omega _{22} =0
\end{equation}
It follows from \eqref{3_17} that the functions $\omega _{pq} $ are linearly dependent. Let us consider the possible variants and identify the solutions that are non-equivalent with respect to the \eqref{1_26_} transformations.

\quad

$\bf{I}$. \quad  Let $ a^{2} -b^{3} = M \ne 0 \rightarrow \omega _{23} = \beta_2\omega _{22} -\alpha_3 \omega _{33},$ \quad

where \quad $(a^3= M\alpha_3, \quad b^2 = M\beta_2).$

If \quad  $\alpha_3\beta_2\geq 0$,\quad using the transformations \eqref{1_26_} one can reduce the matrix $\omega _{\alpha\beta}$ to the diagonal form:
\begin{equation}\label{3_18}
\omega_{\alpha\beta}=\delta_{\alpha\beta}\omega_{\alpha\alpha}.
\end{equation}

If \quad  $\alpha_3\beta_2 < 0$, \quad one can represent the matrix $\omega _{\alpha\beta}$ in the form:
\begin{equation}\label{3_19}
\omega_{\alpha\beta}=\begin{pmatrix}\omega_{11}& 0 & 0 \\ 0 & \omega_{22}  & \omega_{23} \\0 &\omega_{23} & -\omega_{22} \end{pmatrix}.
\end{equation}

$\bf{II}$. \quad \quad  Let $ a^{2} -b^{3} =  0 \rightarrow  b^2\omega _{22} = a^3 \omega _{33}.$

In this case there exist the following non-equivalent with respect to transformations \eqref{1_26_} forms of the matrix $\omega_{pq}$.
\begin{equation}\label{3_20}
\omega_{\alpha\beta}=\begin{pmatrix}\omega_{11}& 0 & 0 \\ 0 & \omega_{22}  & \omega_{23} \\0 &\omega_{23} & \epsilon\omega_{22} \end{pmatrix}.\quad (\epsilon = 0,\pm 1; \quad  b^{2} =\epsilon a^{3}).
\end{equation}
 If in the relations \eqref{3_20} $\epsilon =1$, the matrix $\omega_{\alpha\beta}$ is equivalent to the matrix \eqref{3_18} for the case when \quad $\omega_{22}\omega_{33} > 0.$ If \quad $\epsilon =-1,$\quad we have the variant \eqref{3_19}.
Therefore, there are only three matrices\quad $\omega_{\alpha\beta}$ \quad that are non-equivalent with respect to the \eqref{1_23} transformations:
\begin{equation}\label{3_21}
\omega_{\alpha\beta}=\begin{pmatrix}\omega_{11}& 0 & 0 \\ 0 & \omega_{22}  & 0 \\0 & 0 & \omega_{33} \end{pmatrix},
\end{equation}
\begin{equation}\label{3_22}
\omega_{\alpha\beta}=\begin{pmatrix}\omega_{11}& 0 & 0 \\ 0 & \omega_{22}  & \omega_{23} \\0 &\omega_{23} & -\omega_{22} \end{pmatrix} \quad  (b^{2} =-a^{3}),
\end{equation}
\begin{equation}\label{3_23}
\omega_{\alpha\beta}=\begin{pmatrix}\omega_{11}& 0 & 0 \\ 0 & \omega_{22}  & \omega_{23} \\0 &\omega_{23} & 0 \end{pmatrix} \quad ( b^{2} =0).
\end{equation}

\section{Finding components of the matrix $\omega_{\alpha\beta}$}
Let us find all solutions of the system of equation \eqref{3 14}. We will use the metrics \eqref{3_21} - \eqref{3_23} as the result of the previous classification. For all these metrics, the function $\omega _{11}$ is determined from the last equation of the system \eqref{3 14} and has the form:
\begin{equation}\label{3 20}
\omega _{11} =\exp \left(2\beta +c^{1} \tau \right)
\end{equation}
The functions contain sets of parameters $ c^1, a^p, b^p $. Therefore, below these sets are specified for each of the solutions

\quad

\noindent
$\bf1.$ Let's find the solution for the option \eqref{3_21}. From the equations \eqref{3 14} it follows:
$$
\mathbb{A}_{3} =\mathbb{B}_{2} =0\Rightarrow \mathbb{A}_{2} =a^{2} \omega _{22} ,\qquad \mathbb{B}_{3} =b^{3} \omega _{33}
$$
The non-equivalent solutions of the system \eqref{3 14} have the form ($a^{3} =b^{2} =0$):

\quad

$\bf {a_1}$
\begin{equation} \label{3_24}
 \omega _{22} =\exp a^{2} \tau,\qquad \omega _{33} =\varepsilon \exp b^{3} \tau
\end{equation}

\noindent
$\bf2.$ Consider the options \eqref{3_22}.
$$ \quad \omega _{23} \ne 0 \rightarrow a^{2} =b^{3} ,\quad \omega_{33}^{3} =-\omega _{22} ,\quad b^{2} =-a^{3}.
$$
If $a^{3} =0$, we obtain a special case of the solution of \eqref{3_21}. Therefore, we believe that $a^{3} \ne 0$. From the system \eqref{3 14} it follows:
\begin{equation}\label{3 25}
\omega _{23} =\frac{1}{a^{3} }(\dot{\omega }_{22} -a^{2}\omega_{22})
\end{equation}
\begin{equation}\label{3 26}
\ddot{\omega}_{22} -2a^{2} \dot{\omega}_{22} +({a^2}^2 -\xi{a^{3}}^2)\omega_{22})=0.
\end{equation}
By integrating the system of equations \eqref{3 25}, \eqref{3 25} we find the solutions of the system \eqref{3 14}:

\quad

$\bf {a_2}$ \quad $b^{3} =a^{2},\quad  b^{2} = a^{3}$
\begin{equation} \label{3_27}
\omega _{22} = \omega _{33}=sh(a^{3} \tau )\exp(a^{2} \tau ), \quad \omega _{23} =ch(a^{3} \tau)\exp(a^{2} \tau)  \Rightarrow \varepsilon =-1 \rightarrow
\end{equation}
variable $\tau$ is spatial.

\quad

$\bf {a_3}$ \quad $(b^{3} =a^{2},\quad  b^{2} = -a^{3})$
\begin{equation} \label{3_28}
\omega _{22}=-\omega_{33} =\sin \left(a^{3} \tau \right)\exp \left(a^{2} \tau \right),\quad \omega _{23} =\cos \left(a^{3} \tau \right)\exp \left(a^{2} \tau \right)\Rightarrow \varepsilon =-1\rightarrow
\end{equation}
variable $\tau$ spatial.

\quad

$\bf {a_4}$ \quad $a^{2} =b^{3}, \quad b^{2} = \omega_{33}=0. \Rightarrow \varepsilon =-1\rightarrow $ the variable $\tau $ is spatial.

\quad

$\bf {a_4}$ \quad $a^{2} =b^{3}, \quad b^{2} = \omega_{33}=0. \Rightarrow \varepsilon =-1\rightarrow $ the variable $\tau $ is spatial.

\quad

\noindent
 From the condition \eqref{3 13} we obtain
$$
\left\{\begin{array}{c} {\mathbb{A}_{2} =a^{2} \omega _{22} +a^{3} \omega _{23}, \quad \mathbb{A}_{3} =a^{2} \omega _{23},} \\
{\mathbb{B}_{2} =a^{3} \omega _{23}, \quad \mathbb{B}_{3} =0
}\end{array}\right.
$$
The system of equations \eqref{3 14} will take the form:
\begin{equation} \label{3_29}
\Rightarrow \left\{\begin{array}{c} {\dot{\omega }_{22} =a^{2} \omega _{22} +a^{3} \omega _{23}} \\ {\dot{\omega }_{23} =a^{2} \omega _{23} } \end{array}\right.
\end{equation}
By integrating the system of equations \eqref{3_29}, we obtain:
$$\omega _{23} =p\exp \left(a^{2} \tau \right),\quad \omega _{22} =q\exp \left(a^{2} \tau \right)+\tau a^{3} \omega _{23}. $$
This is equivalent to the solution:
\begin{equation} \label{3_30}
\omega _{22} = a^{3} \tau\exp(a^{2} \tau), \quad\omega _{23} =\exp (a^{2} \tau ),\quad \omega _{33} = 0.
\end{equation}
Let us give all the solutions of the autonomous system.

\quad

\noindent
{\bf 1.} \quad $b^{2} =a^{3}.$
\begin{equation}\label{3_31}
\omega_{\alpha\beta}=\begin{pmatrix}\exp \left(2\beta +c^{1} \tau \right) & 0 & 0 \\ 0 & \exp( a^{2} \tau)  & 0 \\0 & 0 & \varepsilon \exp (b^{3} \tau) \end{pmatrix} ,
\end{equation}

\quad

\noindent
{\bf 2.} \quad $b^{2} =-a^{3}.$
\begin{equation}\label{3_32}
\omega_{\alpha\beta}=\begin{pmatrix}\exp (2\beta +c^{1} \tau )& 0 & 0 \\ 0 & sh(a^{3} \tau )\exp (a^{2} \tau ) & ch(a^{3} \tau )\exp (a^{2} \tau ) \\0 &ch(a^{3} \tau )\exp (a^{2} \tau ) & -sh(a^{3} \tau )\exp (a^{2} \tau ) \end{pmatrix},
\end{equation}

\noindent
{\bf 3.} \quad $b^{2} =-a^{3}.$
\quad
\begin{equation}\label{3_33}
\omega_{\alpha\beta}=\begin{pmatrix}\exp (2\beta +c^{1} \tau )& 0 & 0 \\ 0 & \sin(a^{3} \tau )\exp (a^{2} \tau ) & \cos(a^{3} \tau )\exp (a^{2} \tau ) \\0 &\cos(a^{3} \tau )\exp (a^{2} \tau ) & -\sin(a^{3} \tau )\exp (a^{2} \tau ) \end{pmatrix},
\end{equation}

\quad

\noindent
{\bf 4.} \quad $b^{2} =0.$
\begin{equation}\label{3_34}
\omega_{\alpha\beta}=\begin{pmatrix}\exp (2\beta +c^{1} \tau )& 0 & 0 \\ 0 & \tau a^{3}\exp(a^{2} \tau)  & \exp(a^{2} \tau) \\0 &\exp(a^{2} \tau) & 0 \end{pmatrix},
\end{equation}

\quad

\section{Components of the metric tensor $\eta_{\alpha\beta}$}
The components of the metric tensor $\eta _{\alpha \beta } $ are expressed through the functions $\omega _{\alpha \beta }$ \eqref{1_31_} - \eqref{1_33_} using the relation \eqref{1_36_}:
\begin{equation} \label{4.1}
\eta _{\alpha \beta } =\omega _{\alpha \beta } \exp \left(\gamma -\beta \right).
\end{equation}
The functions $\beta $, $\gamma $ obey the system of equations \eqref{1_32_}, \eqref{1_33_}. From \eqref{1_33_} it follows:
\begin{equation}\label{4.2}
\left\{\begin{array}{c} {\kappa _{\alpha }^{\alpha } =3\dot{\gamma }-\dot{\beta }+C_{\alpha }^{\alpha } \Rightarrow} \qquad\qquad \qquad \qquad \qquad  \\{ \left(\kappa _{\alpha }^{\alpha } \right)^{2} =\left(3\dot{\gamma }-\dot{\beta }\right)^{2} +2\left(3\dot{\gamma }-\dot{\beta }\right)C_{\alpha }^{\alpha } +\left(C_{\alpha }^{\alpha } \right)^{2},} \qquad
 \\ {\kappa _{\beta }^{\alpha } \kappa _{\alpha }^{\beta } =3\dot{\beta }^{2} +3\dot{\gamma }^{2} -2\dot{\gamma }\dot{\beta }+4C_{1}^{1} \dot{\beta }-2\left(\dot{\gamma }-\dot{\beta }\right)C_{\alpha }^{\alpha } +C_{\beta }^{\alpha } C_{\alpha }^{\beta }.} \end{array}\right.
\end{equation}
Using the \eqref{4.2} relations, we reduce the equation \eqref{1_33_} to the form:
\begin{equation} \label{4.3}
2\left(\ddot{\gamma }-\ddot{\beta }\right)+\left(\dot{\beta }+\dot{\gamma }\right)^{2} -4\dot{\gamma }^{2} +2\dot{\beta }c^{1} -2\dot{\gamma }C_\alpha^\alpha +a^{3}b^{2} -a^{2} b^{3} -c^{1} C^p_p=0.
\end{equation}
For all matrices $\omega _{\alpha \beta } $ found in the previous section, the sets of values $\eta _{11} , \ell ^{2} $ have the same form:
\begin{equation} \label{4_5_}
\eta _{11} =\exp \left(\gamma +\beta +C_{1}^{1} \tau \right),\qquad \ell ^{2} =\exp \left(3\gamma -\beta +C_{\alpha }^{\alpha } \right),
\end{equation}
and differ only by the values of the parameters $a^{p} , b^{p} $, which are specified for each of the above solutions of the autonomous system.
Using \eqref{4_5_}, let us represent the equations of \eqref{1_32_} in the form:
\begin{equation} \label{4_6}
\ddot{\beta}=\tilde{p}\exp \left(\gamma +\beta +C_{1}^{1} \tau \right)
\end{equation}
\begin{equation} \label{4_7}
\ddot{\gamma}=q\exp \left(3\gamma -\beta +C_{\alpha }^{\alpha } \tau \right),
\end{equation}
where
\begin{equation} \label{4_8}
\tilde{p}=\frac{\kappa \alpha ^{2} }{2\pi }, \quad q=4\xi|\lambda|.
\end{equation}
Let us show that the equation \eqref{4.3} is joint with the system of equations \eqref{4_6}, \eqref{4_7}. To do this, we substitute the functions $\ddot{\beta }, \ddot{\gamma }$ from the system \eqref{4_6}, \eqref{4_7} into equation \eqref{4.3}. Then we differentiate the resulting equation:
\[2q\exp \left(3\gamma -\beta +C_{\alpha }^{\alpha } \tau \right)-2p\exp \left(\gamma +\beta +C_{1}^{1} \tau \right)+\left(\dot{\beta }+\dot{\gamma }\right)^{2}-\]
\begin{equation} \label{4_9 }
-4\dot{\gamma }^{2} +2\dot{\beta }C_{1}^{1} -2\dot{\gamma }C_{\alpha }^{\alpha } +a^{3} b^{2} -a^{2} b^{3} -C_{1}^{1} \left(a^{2} +b^{3}\right)=0
\end{equation}
by the $\tau$ variable
We replace the functions \quad $\ddot{\gamma }, \quad\ddot{\beta }$ \quad by the right parts of the equations \eqref{4_6}, \eqref{4_7}. As a result, we obtain zero.
Let us give all the obtained solutions:

\quad

$\bf 1$ \quad $(a^3 = b^2 =0)$
\begin{equation} \label{4_10_}
ds^{2} =-\varepsilon\exp(3\gamma -\beta +(a^2 + b^3 + c^1)\tau) {d\tau}^{2} +
 \exp (\gamma + \beta +c^{1}\tau){du^1}^{2}
+\exp(\gamma -\beta)(\exp (a^{2}\tau) {du^2}^{2} +\end{equation}
$$\varepsilon\exp (b^{3} \tau) {du^3}^{2}).$$
\quad

\quad

$\bf 2$ \quad $(b^2 = a^3,\quad b^3 =a^2)$
\begin{equation}\label{4 11}
ds^{2} =\exp(3\gamma + \beta +(a^2 + b^3 + c^1)\tau) {d\tau}^{2} +
\exp (\gamma + \beta +c^{1}\tau){du^1}^{2}+\exp(\gamma -\beta + a^{2}\tau) ({du^2}^{2}\sinh(a^{3}\tau) +\end{equation} $$
 {du^3}^{2}\sinh(a^{3} \tau) +2du^{2}du^{3}\cosh(a^3\tau)).$$
\quad
$\bf 3$ \quad $(b^2 = -a^3, \quad b^3 =a^2)$
\begin{equation}\label{4 12}
ds^{2} =\exp(3\gamma -\beta +(a^2 + b^3 + c^1)\tau) {d\tau}^{2} +
\exp (\gamma + \beta +c^{1}\tau){du^1}^{2}+ \exp(\gamma -\beta + a^{2}\tau) ({du^2}^{2}\sin(a^{3}\tau) +
\end{equation}
$$+ {du^3}^{2}\sin(a^{3} \tau) +2du^{2}du^{3}\cos(a^3\tau)).$$

\quad

$\bf 4$ \quad $(b^2 =0,\quad b^3 =a^2)$
\begin{equation}\label{4 13}
ds^{2} =\exp(3\gamma -\beta +(a^2 + b^3 + c^1)\tau) {d\tau}^{2}+
\exp (\gamma + \beta +c^{1}\tau){du^1}^{2}+
\end{equation}
$$+\exp(\gamma -\beta + a^{2}\tau) (\tau{du^2}^{2} +2du^{2}du^{3})
$$

To find electrovacuum solution of the Einstein-Maxwell equations one has to integrate the system of equations \eqref{4.3}, \eqref{4_6}, \eqref{4_7}.

\section{Solution of the system of equations \eqref{4.3}-\eqref{4_7}}

Let's introduce functions \quad $\varphi, \rho$ \quad and parameter $p$ that satisfy the conditions:
\begin{equation}\label{1}
\left\{\begin{array}{c} {\varphi =4\gamma +\left(2c^{1} +a^{2} +b^{3} \right)\tau +\ln{(\frac{2\kappa\alpha^{2}|\lambda |}{\pi})}} \qquad \\{}
\\ {\rho=\gamma +\beta +c^{1} \tau +\ln\left(\frac{\kappa\alpha^{2}}{2\pi } \right)} \qquad \qquad  \qquad  \qquad
\\{}
\\ \epsilon p^{2}={a^{3} b^{2} +\frac{1}{4} \left(a^{2} -b^{3} \right)^{2} \quad \epsilon \pm 1 \qquad \qquad \qquad} \end{array}\right.
\end{equation}
Then the system of equations \eqref{4_6}- \eqref{4_8} can be represented in the form:
\begin{equation} \label{2}
\left\{\begin{array}{c} {\ddot{\varphi }=4\xi\exp \left(\varphi -\rho \right)\qquad \qquad \qquad \qquad \qquad \qquad } \\
 {\ddot{\rho }=\exp \rho +\xi\exp \left(\varphi -\rho \right)\qquad \qquad \qquad \qquad \qquad } \\ {\ddot{\varphi }-2\ddot{\rho }+\dot{\rho }^{2} -\frac{1}{4} \dot{\varphi }^{2} +\epsilon p^{2} =0\qquad \qquad \qquad \qquad \qquad } \end{array}\right.
 \end{equation}
From the first equation of the system \eqref{2} follows:
\begin{equation} \label{4_26_}
\rho =\ln{4}-\ln{\xi\ddot{\varphi}} +\varphi
\end{equation}
Let us substitute \eqref{4_26_} into the second equation of the system \eqref{2}. The result is:
\begin{equation} \label{4_27_}
\dot{\chi }=-\frac{\exp \varphi }{\varphi ^{\left(2\right)} }
\end{equation}
where
$$\qquad \chi = \left(\frac{\varphi^{\left('''\right)} }{\varphi^{\left(''\right)} } \right)^{(')}.$$
Using the equations \eqref{4_26_}, \eqref{4_27_}, we reduce the last equation of the system to the form:
\begin{equation} \label{4_28_}
\left(\dot{\varphi }-\chi \right)^{2} =\Phi, \qquad \Phi =\ddot{\varphi }+\frac{2\exp \varphi }{\ddot{\varphi }} +\frac{1}{4} \dot{\varphi }^{2} -\epsilon \rho ^{2}
\end{equation}
Thus, we obtained the only equation to which the function $\varphi$ obeys. The function $\varphi$ defines the functions $\gamma,\beta$:
\[\gamma =\frac{\varphi}{2} -\frac{1}{4}( \ln\frac{\xi q \tilde{p}\ddot{\varphi}}{4}+(2c^{1} +a^{2} +b^{3} )\tau )\]
\[\beta =\frac{\varphi}{2} -\frac{3}{4} \ln{\xi\ddot{\varphi}}+\frac{1}{4}(a^{2} +b^{3} - 2c^{1})\tau  + \ln q(\frac{4}{\tilde{p}})^3 \]
 The equation \eqref{4_28_} is nonlinear, third order. The senior derivative can be explicitly expressed through the other function:
\begin{equation}\label{4_29_}
\chi =\dot{\varphi }+\xi _{1} \sqrt{\Phi } \qquad \xi _{1} =\pm 1,
\end{equation}

The order of the equation \eqref{4_29_} can be lowered by introduction a new variable $u=\varphi$ and a new function $W = (\dot{\varphi})^2$. A general solution for the nonzero cosmological term has not been found yet.

\subsection{Partial solutions for the case $\lambda \ne 0$ }

The integration problem of an ordinary differential equation belongs to the separate branch of mathematics and it is not an object for research for the classification  of exact solutions of the gravitational field equations. Let us specify some partial solutions that can be found in analytic form under additional conditions. We choose the parameters that enter the equations \eqref{4.3} - \eqref{4_7} as follows:

 \begin{equation}\label{6.1}
 \xi =1, \quad\alpha = 2\sqrt{\frac{2|\lambda|\pi}{\kappa}},
 \end{equation}
 $$ (a^2 - b^3)^2 +4a^3 b^2 =0 $$
Let us show that in this case there exists a partial solution of the equations \eqref{4.3} -\eqref{4.7}  of the form: the system of
$$
\gamma =\beta -(\frac{a^2 + b^3}{2})\tau.
$$
The function $\beta$ obeys the equations:

\begin{equation}\label{6.2}
\ddot{\beta }=\exp(2\beta + \frac{1}{2}(2c^1-a^2-b^2)\tau + \ln{\frac{\kappa\alpha^2}{2\pi}}).
\end{equation}
Let us denote:
$$
y= \beta + \frac{1}{4}(2c^1-a^2-b^2)\tau + \ln{\frac{\kappa\alpha^2}{2\pi}}.
$$
Then the solutions of the equations \eqref{4_7}, \eqref{4_7} will have the form:

\begin{equation}\label{6.3}
1. \quad y =\ln{\frac{1}{\sinh\tau }}, \quad   2.\quad y =\ln{\frac{1}{\sin \tau }}, \quad  3.\quad y =\ln{\left(\frac{1}{\tau} \right)}.
\end{equation}
Functions $\beta,\gamma$ can be represent in the form:
$$
\beta = y - \frac{1}{4}(2c^1-a^2-b^2)\tau - \frac{1}{2}\ln{\frac{\kappa\alpha^2}{2\pi}},
$$
$$
\gamma = y - \frac{1}{4}(2c^1+a^2+b^2)\tau - \frac{1}{2}\ln{\frac{\kappa\alpha^2}{2\pi}},
$$
 Using the metrics \eqref{4_10_}, \eqref{4 13} and the relations \eqref{6.1}, \eqref{6.3}, we obtain the following list of solutions:

\quad

$\bf{1}$
\begin{equation} \label{6.4}
ds^{2} =\frac{ \exp2y }{4|\lambda| } ({-\varepsilon{d\tau}^2}+{du^1}^{2}) + ({{du}^{2}}^2 +\varepsilon{du^3}^2)\exp a^2\tau
\end{equation}

\quad

$\bf{2}$

\begin{equation} \label{6.5}
ds^{2} =\frac{ \exp2y }{4|\lambda| } ({{d\tau}^2}+{du^1}^{2}) + (({{du}^{2}}^2 +{du^3}^2 du^{3})\sinh a^3\tau +
\end{equation}
$$+2 {du^3}du^{3}\cosh a^3\tau)\exp a^2\tau ;$$
\quad

$\bf{3}$
\begin{equation} \label{6.5}
ds^{2} =\frac{ \exp2y }{4|\lambda| } ({{d\tau}^2}+{du^1}^{2}) + (({{du}^{2}}^2 - {du^3}^2 du^{3})\sin a^3\tau +
\end{equation}
$$+2 {du^3}du^{3}\cos a^3\tau)\exp a^2\tau ;$$

$\bf{4}$

\begin{equation} \label{6.7}
ds^{2} =\frac{ \exp2y }{4|\lambda| } ({{d\tau}^2}+{du^1}^{2}) +
\end{equation}
$$+(\tau{{du}^{2}}^2 +2{du}^2 du^{3})\exp a^2\tau.$$
The function $y$ is given by the relations \eqref{6.3}.

\subsection{Complete classification for the case $\lambda =0$ }

 Equations \eqref{4_6}, \eqref{4_9 } for the case, when $\gamma =0$ have the form:
\begin{equation} \label{5.1}
\ddot{\beta }=q\exp \left(\beta +c^{1} \tau \right) \quad \left(q=\frac{\kappa \alpha ^{2} }{2\pi } \right),
\end{equation}
\begin{equation} \label{5.2}
2\ddot{\beta }=\left(\dot{\beta }+c^{1} \right)^{2} +4\epsilon p^{2},
\end{equation}
where $\quad \epsilon=0\pm 1,\quad 4\epsilon p^{2} =a^{3} b^{2} +\frac{1}{4} \left(a^{2} -b^{3} \right)^{2} -\left(c^{1} +\frac{1}{2} \left(a^{2} +b^{3} \right)\right)^{2}$,
\begin{equation} \label{5.3}
c^{1} =-\frac{1}{2} \left(a^{2} +b^{3} \right)+\xi_1 \sqrt{\frac{1}{4} \left(a^{2} -b^{3} \right)^{2} +a^{3} b^{2} -4\epsilon p^{2}},\end{equation}\quad
$\xi_1 =\pm 1$ .
From equation \eqref{5.2} it follows:
\begin{equation} \label{5.4}
\int \frac{d\left(\dot{\beta }+c^{1} \right)}{\left(\dot{\beta }+c^{1} \right)^{2} +4\epsilon p^{2} } =\frac{\tau }{2}.
\end{equation}
Find the function $\beta+c^{1}$ from \eqref{5.4} and substitute it into \eqref{5.2}. The solution can be represented as:
\begin{equation}\label{5.5}
\beta +c^1\tau = \ln k -2\ln y.
\end{equation}
Here:
$$
k= \quad \left\{\begin{array}{c} 4\pi p^2 /\kappa\alpha^2 \quad (\epsilon^2 =1), \\ {\quad} \\
 4\pi/\kappa\alpha^2 \quad (\epsilon =0). \end{array}\right.
$$
 $y$, depending on the value of $\epsilon$ has the form:

\quad

$\bf{1}$
\begin{equation}\label{5.6}
\epsilon =1,\quad y = \sin{p\tau},
\end{equation}

\quad

$\bf{2}$
\begin{equation}\label{5.7}
\epsilon =-1, \quad y = \sinh{p\tau},
\end{equation}

$\bf{3}$
\begin{equation}\label{5.8}
\epsilon =0, \quad y = \tau.
\end{equation}

Thus, each of the four metrics represented by the formulas \eqref{4_10_}, \eqref{4 13} defines three solutions of the electrovacuum Einstein equations. A complete classification of non-isotropic electrovacuum spaces with a zero cosmological term, invariant with respect to the action of the three-parameter group of motions, is obtained. In contrast to the case with a non-zero cosmological term, when the classification is carried out with accuracy to the solution of one ordinary differential equation, in this case all electrovacuum solutions are given in elementary functions.

 Using metrics \eqref{4_10_}, \eqref{4 13}, relations \eqref{5.6}-\eqref{5.8} and  re-defining constant parameters \quad ($a, b, p  =const$), we obtain the following list of solutions:

\quad

{\bf 1} metric \eqref{4_10_},  $a^3 = b^2 =0$. 
$$\quad c^1 = \xi_1b -\frac{a^2 +b^3}{2}, \quad a = \frac{a^2 - b^3}{2}, \quad b^2= \frac{(a^2 - b^3)^2}{4} -4\epsilon p^2.$$

\quad

1) $\quad \epsilon =1$
\begin{equation} \label{5 9}
ds^{2} = -\varepsilon\frac{\sin^2{p\tau}}{k} \exp (2b\xi_1\tau) ( d\tau)^{2} + \frac{k}{\sin^2{p\tau}}{du^1}^2
+ \frac{\sin^2{p\tau}}{k}\exp( b\xi_1\tau) (\exp a\tau(du^2)^2 +\varepsilon \exp( -a\tau)(du^3)^2),
\end{equation}

\quad

2) $\quad \epsilon =-1$.
\begin{equation} \label{5 9}
ds^{2} = -\varepsilon\frac{\sinh^2{p\tau}}{k} \exp (2b\xi_1\tau) ( d\tau)^{2} + \frac{k}{\sinh^2{p\tau}}{du^1}^2
+ \frac{\sinh^2{p\tau}}{k}\exp( b\xi_1\tau) (\exp a\tau(du^2)^2 + ,
\end{equation}
$$\varepsilon \exp( -a\tau)(du^3)^2)$$

3) $\quad (\epsilon =0, \quad a^2 +b^3 = 2a \quad a^2 -b^3 = 2b).$
\begin{equation} \label{5 9}
ds^{2} =-\varepsilon \frac{\tau^2{d\tau}^{2}}{k} \exp (2\xi_1 b\tau)   + \frac{k}{\tau^2}{du^1}^2
+ \frac{\tau^2}{k}((du^2)^2\exp b(\xi_1 +1)\tau +\varepsilon (du^3)^2\exp b(\xi_1 -1)\tau)
\end{equation}

{\bf 2} \quad Metric \eqref{4 11}. \quad $b^2 = a^3,\quad b^3 =a^2$.

\quad

4) \quad $\epsilon = 1, \quad c^1=2\xi_1 p\sinh a, \quad a^3 =2p\cosh a.$

\quad
\begin{equation} \label{5 12}
ds^{2} =\frac{\sin^2(p\tau)}{k}(\exp (4\xi_1 p \tau \sinh a){d\tau}^{2}+ \frac{k}{\sin^2(p\tau)}{du^1}^2
\frac{\sin^2(p\tau)}{k} \exp (2\xi_1 p \tau \sinh a)(({du^2}^2 +
\end{equation}
$$
{du^3}^2)\sinh(2p\tau\cosh a)++2du^2du^3\cosh(2p\tau\cosh a))
$$

\quad

5) \quad $\epsilon = -1, \quad c^1=2\xi_1 p\sin a, \quad a^3 =2p\cos a.$
\quad
\begin{equation} \label{5 12}
ds^{2} =\frac{\sinh^2(p\tau)}{k}(\exp (4\xi_1 p \tau \sin a){d\tau}^{2}+ \frac{k}{\sinh^2(p\tau)}{du^1}^2
\frac{\sinh^2(p\tau)}{k} \exp (2\xi_1 p \tau \sin a)(({du^2}^2 -
{du^3}^2)\sinh(2p\tau\cos a)+\end{equation}
$$+2du^2du^3\cosh(2p\tau\cos a))
$$

6) \quad $\epsilon = 0, \quad c^1=\xi_1 a^2 -b,\quad a^3 = b. $
\begin{equation} \label{5 11}
ds^{2} =\frac{\tau^2 d\tau ^{2} }{k} \exp (2\xi_1\tau b) + \frac{k}{\tau^2}{du^1}^2
+ \frac{\tau^2 }{k}\exp(\xi_1\tau b)(({du^2}^2 +
{du^3}^2)\sinh(b\tau) +2du^2du^3\cosh(b\tau))\end{equation}

\quad

{\bf 3} \quad Metric \eqref{4 12}. \quad $b^2 = -a^3,\quad b^3 =a^2$.

\quad

7)\quad $\epsilon = -1, \quad c^1=2\xi_1 p\cos a, \quad a^3 =2p\sin a.$
\quad
\begin{equation} \label{5 12}
ds^{2} =\frac{\sinh^2(p\tau)}{k}(\exp (4\xi_1 p \tau \cos a){d\tau}^{2}+ \frac{k}{\sinh^2(p\tau)}{du^1}^2
frac{\sinh^2(p\tau)}{k} \exp (2\xi_1 p \tau \cos a)(({du^2}^2 -\end{equation}
$$
-{du^3}^2)\sin(2p\tau\sin a)+2du^2du^3\cos(2p\tau\sin a)).
$$

\quad
{\bf 4} \quad Metric \eqref{4 13}. \quad $b^2 = 0,\quad b^3 =a^2$.

\quad

8) \quad $\epsilon = -1, \quad c^1=2\xi_1 p - a^2.$
\begin{equation} \label{5 14}
ds^{2} =\frac{\sinh^2(p \tau)}{k} \exp (4\xi_1 p\tau ) ( d\tau )^{2} + \frac{k}{\sinh^2(p \tau)}{du^1}^2
+\frac{\sinh^2(p \tau)}{k} \exp (2\xi_1 p\tau ) (\tau {du^2}^2
+2du^2du^3)
\end{equation}

\quad

9) \quad $\epsilon = 0, \quad c^1= - a^2.$
\begin{equation} \label{5 14}
ds^{2} =\frac{\tau^2}{k}(\tau d\tau)^{2} + \frac{k}{\tau^2}{du^1})^2+
\frac{\tau^2}{k} (\tau {du^2}^2
+2du^2du^3)\end{equation}

\quad

\section{Conclusion}
The problem of classification of electrovacuum spaces with three-parameter Abelian groups of motions acting on non-isotropic hypersurfaces has been considered in the paper. In the general case, the problem was reduced to integrating of a single nonlinear ordinary differential equation. Thus, the considered classification is completed, since when applying symmetry theory methods to systems of differential equations such a result is the ultimate aim.  In this paper, several partial solutions of this equation are also found in analytic form. In the case of a zero cosmological term, the full solution of the classification problem is obtained in explicit form. Obtaining the complete decision of the considered classification problem, on the one hand, means the completion of the complete classification of stackel spaces of  electrovacuum.  On the other hand, the first example of classification for Petrov spaces of the electrovacuum is obtained, since the considered stackel spaces with three-parameter groups of motions belong to the Bianchi of type I spacetime.

\quad

Data availability
No data was used for the research described in the article.

Acknowledgments
The work is supported by Russian Science Foundation, project number N 23-21-00275.

https://rscf.ru/en/project/23-21-00275/.


\begin{thebibliography}{99}

\bibitem[Stephani et al. (2003)]{1}
Stephani H., Kramer D., Mac Callum  M., Hoenselaers C., and Herlt E.
Exact Solutions of Einstein's Field Equations. Second Edition.{\em Cambridge University Press. Cambridge.} {\bf 2003} (732 pp.) ISBN 0521461367.doi: https://doi.org/10.1017/CBO9780511535185

\bibitem[Petrov (1951)]{2} Petrov A. Z. Einstein Spaces, {\bf Oxford}, {\bf 1969}. ( Russian original published by Nauka, Moscow, 1951).

\bibitem[Errehymy et al. (2021)]{1a}
Errehymy, A., Khedif, Y. and Daoud, M. Anisotropic compact stars via embedding approach in general relativity: new physical insights of stellar configurations. {\em  Eur. Phys. J.} {\bf 2021} C 81, 266 . https://doi.org/10.1140/epjc/s10052-021-09062-3

\bibitem[Nikolaev et al. (2020)]{1b}
Nikolaev, A.V., Maharaj, S.D. Embedding with Vaidya geometry. {\em Eur. Phys. J.} {\bf 2020} C 80, 648. https://doi.org/10.1140/epjc/s10052-020-8231-0

\bibitem[Shapovalov (1978)]{3}
 Shapovalov V.N., Symmetry and separation of variables in the Hamilton-Jacobi equation. {\em Sov. Phys.J.}.  {\bf 1978}, {\em 21}, 1124-1132pp.      doi: 10.1007/BF00894560;

\bibitem[Shapovalov (1979)]{4}
 Shapovalov V.N., Stackel`s spaces. {\em Sib. Math. J.} {\bf 1979}, {\em 20}, (1117-1130pp.), doi: org/10.1007/BF00971844;

\bibitem[Shapovalov (1975)]{5} Shapovalov V.N., Symmetry of motion equations of free particle  in riemannian space. {\em Sov. Phys.J.} {\bf 1975}, {\em 18}, (1650-1654pp.), doi.org/10.1007/BF00892779;

\bibitem[Shapovalov (1978)]{6}
Shapovalov V.N., Symmetry and separation of variables in a linear second-order differential equation. I, II. {\em Sov. Phys.J.}, {\bf 1978} {\em21}, (645-650, 693-695 pp.) doi.org/10.1007/BF00890983;

\bibitem[Shapovalov (1975)]{7} Shapovalov V.N., Symmetry of Dirac-Fock equation
{\em Soviet Physics Journal} 18 (6), 797-802. doi:org/10.1007/BF00891156

\bibitem[Obukhov (2020)]{8}
Obukhov V.V. Hamilton-Jacobi equation for a charged test particle in the Stackel space of type (2.0). {\em Symmetry} {\bf 2020}, 12(8), 1289; doi.org/10.3390/sym12081289;

\bibitem[Bagrov et al. (1973)]{9}
Bagrov, A. G. Meshkov, V. N. Shapovalov, A. V. Shapovalov. Separation of variables in the Klein-Gordon equations I. {\em Sov. Phys.J} {\bf1973}, {\em16}, (1533-1538 pp.), doi: 10.1007/BF00889957 ;

\bibitem[Bagrov et al. (1973)]{10}
 Bagrov, A. G. Meshkov, V. N. Shapovalov, A. V. Shapovalov. Separation of variables in the Klein-Gordon equations II. {\em Sov. Phys.J} {\bf1973}, {\em16}, (1659-1665pp.), doi: org/10.1007/BF00893656;

\bibitem[Bagrov et al. (1974)]{11}
Bagrov, V. G. Meshkov,A. G., Shapovalov, V. N. Shapovalov A. V.. Separation of variables in the Klein-Gordon equations III. {\em Sov. Phys. J.} {\bf 1974}, {\em17}, (812-815 pp.) doi:org/10.1007/BF00890216;

\bibitem[Shapovalov et al. (1973)]{12}
Shapovalov V.N., Eckle G.G., Separation of Variables in the Dirac Equation. {\em Sov. Phys. J.} {\bf 1973}, {\em16}, (818-823pp.), doi: org/10.4213/tmf1093;

\bibitem[Schwarzschild (1916)]{13}
Schwarzschild, K. Uber das Gravitationsfeld eines Masenpunktes nach der
Einsteinschen Theorie.{\em Sitz. Preuss. Akad. Wiss.}, {\bf 1916},Seite 189-196.

\bibitem[Schwarzschild (1916)]{14}
Schwarzschild, K.  Uber das Gravitationsfeldeiner Kugel aus inkompressibler Flussigkeit nach der Einsteinschen Theorie. {\em Sitz. Preuss. Akad. Wiss.}, {\bf 1916}, 424-434.

\bibitem[Kerr (1963)]{15}
Kerr, R.P.. Gravitational field of a spinning mass as an example of algebraically
special metrics. {\em Phys. Rev. Lett.} {\bf 1963}, 11, 237, doi.org/10.1103/PhysRevLett.11.237;

\bibitem[Newman et al. (1963)]{16}
Newman,E. Tamburino L. and Unti T.,Empty space generalization of the Schwarzschild metric, {\em J. Math. Phys.} {\bf1963}, 915,doi:10.1063/1.1704018

\bibitem[Reissner (1916)]{17}
Reissner, H.. Uber die Eigengravitation des elektrischen Feldes nach der Einsteinschen Theorie., {\em Annalen der Physik}, {\bf 1916}, 355, 9, (106-120), doi:org/10.1002/andp.19163550905

\bibitem[Friedmann (1924)]{18}
Friedmann, A. Uber die Moglichkeit einer Welt mit konstanter negativer Krummung des Raumes. {\em Z. Physik}, 21, 326-332 {1924}. doi.org/10.1007/BF01328280

\bibitem[Fomin et al. (2017)]{19} Fomin I.V., Chervon S. V. Exact and approximate solutions in the Friedmann cosmology. {\em Russ. Phys. J.},{\bf 2017} {\em 60}, {\em issue 30}, (427-440pp.), doi.org/10.1007/s11182-017-1091;

\bibitem[Odintsov et al. (2023)]{20}
 Odintsov, S.D.; Oikonomou,V.K.; Giannakoudi, I.; Fronimos, F.P.;Lymperiadou, E.C. Recent Advances in Inflation. {\em Symmetry}, {\bf2023}, 15, 1701. https://doi.org/10.3390/
sym15091701

\bibitem[Nojiri et al. (2017)]{21}
  Nojiri,S., Odintsov S.D., and Oikonomou V.K. Modified gravity theories on a nutshell: Inflation, bounce and late-time evolution.  {\em Phys. Rept.}, {\bf2017}, (692pp.), doi:10.1016/j.physrep.2017.06.001;

\bibitem[Bamba et al. (2012)]{22}
Bamba  K., S.Capozziello S., Nojiri  S. and Odintsov S.D.,
  Dark energy cosmology: the equivalent description via different theoretical models and cosmography tests. {\em Astrophys. Space Sci.},{\bf 2012}, {\em342}, (155pp.), doi: 10.1007/s10509-012-1181-8;

\bibitem[Capozziello et al. (2012)]{23}
Capozziello S., De Laurentis M., Odintsov D.Hamiltonian dynamics and Noether symmetries in extended gravity cosmology.  {\em Eur.Phys.J.} {\bf 2012}, {\em C72}, 2068 (22 pp.), doi: 10.1140/epjc/s10052-012-2068-0;

\bibitem[Osetrin et al. (2016)]{24}
Osetrin K.E., Filippov A.E., Osetrin E.R. The spacetime models with dust matter that admit separation of variables in Hamilton-Jacobi equations of a test particle. {\em Modern Physics Letters A,} {\bf 2016}, {\em 31}, 1650027 (6pp.), doi:org/10.1142/S0217732316500279;

\bibitem[Maharaj et al. (2017)]{25}
Maharaj S.D., Goswami R., Chervon S. V. and  Nikolaev A. V. Exact solutions for scalar field cosmology in f(R) gravity. {\em Modern Physics Letters AVol.} {\bf 2017}, {\em32}, {\em No. 30}, 1750164 (18pp.), doi.org/10.1142/S0217732317501644;

\bibitem[Vasudevan et al. (2005)]{26}
Vasudevan M., K.A. Stevens And D.N. Page, Separability Of The Hamilton-Jacobi And Klein-Gordon Equations In Kerr-De Sitter Metrics. {\em Class. And Quant. Grav.} {\bf 2005}, {\em22}, (339-352pp.), doi: 10.1088/0264-9381/22/2/007;


\bibitem[Frolov et al. (2018)]{27}
 Valeri P. Frolov, Kyoto U., Pavel Krtous , David Kubiznak. Separation of variables in maxwell equations in Plebanski-Demianski spacetime. {\em Phys.Rev.} {\bf 2018}, {\em D 97}, {\em No.10}, 101701 (6 pp.), doi:10.1103/PhysRevD.97.101701;

\bibitem[Chong et al. (2005)]{28}
Chong Z.W., Gibbons G,W. and Pope C.N. Separability and Killing tensors in Kerr-Taub-Nut-De Sitter metrics in higher dimensions. {\em Phys. Lett. B.} {\bf 2005}, {\em 609}, (124-132pp.), doi: 10.1016/j.physletb.2004.07.066;


\bibitem[Shapovalov et al. (1996)]{29}
Shapovalov, A.V.; Shirokov, I.V. Noncommutative Integration Method For Linear Partial Differential Equations. Functional Algebras And Dimensional Reduction. {\em  Theoret. And Math. Phys.} {\bf 1996}, {\em 106:1}, (1-10 pp).

\bibitem[Osetrin et al. (2020)]{30}
 K. E. Osetrin and Epp V. Y. and Chervon S. V. Propagation of light and retarded time of radiation in a strong gravitational wave, {\em Annals of Physics},
{\bf 2024}, {\em 462}, (169619), doi.org/10.1016/j.aop.2024.169619	

\bibitem[Osetrin et al. (2020)]{31}
Osetrin K., Filippov A., and Osetrin E. Wave-like spatially homogeneous models of Stackel spacetimes (2.1) type in the scalar-tensor theory of gravity.{\em  Modern Physics Letters.} {\bf2020}, AVol., 35, No. 33, 2050275. doi.org/10.1142/S0217732320502752

\bibitem[Carter (1968)]{32}
Carter B. New family of Einstein spaces. {\em Phys.Lett.}
{\bf 1968}, {\em A.25}, {\em No 9} (399-400pp.), doi.org/10.1016/0375-9601(68)90240-5;

\bibitem[Miller et al. (2013)]{33}
Miller Jr. W., Post S., Winternitz P., Classical Aad Qqantum superintegrability with applications. {\em J. Phys. A: Math. Theor.} {\bf 2013}, {\em 46}, 423001, (97 pp.), doi: 10.1088/1751-8113/46/42/423001;

\bibitem[Boyer et al. (1981)]{34}
Boyer C.P., Kalnins E.G., Miller W. Separation of variables in Einstein spaces. I. Two ignorable and one null coordinate. {\em J.Phys. Math. Gen.} {\bf 1981}, {\em 14},{\em No 7}, (1675-1684pp.), doi: org/10.1088/0305-4470/14/7/023;

\bibitem[Bagrov et al. (1983)]{35}
Bagrov V.G., Obukhov V.V. Classes of exact solutions of the Einstein-Maxwell equations. {\em Ann. der Phys.}. {\bf 1983}, {\em B 40}, {\em H 4/5}, (181-188 pp.), doi:10.1002/andp.19834950402;

\bibitem[Bagrov et al. (1986)]{36}
Bagrov V.G., Obukhov V.V.,Shapovalov A.V. Special Stackel electrovac spacetimes. {\em Pramana J. Phys.}. {\bf 1986}, {\em 26} {\em No 2}, (93-108pp.),doi:org/10.1007/BF02847629;

\bibitem[Bagrov et al. (1994)]{37}
V.G. Bagrov, V.V. Obukhov. "Separation of variables for the Dirac square equation". {\em International Journal of Modern Physics D. 3}, {\bf 04}, {\bf 12}, 739-746 (1994). doi: 10.1142/S021827189400085

\bibitem[Magazev (2012)]{38}
A.A.Magazev, "Integrating Klein-Gordon-Fock equations in an extremal electromagnetic
field on Lie groups". {\em Theor.and Math.Phys.} {\bf 2012} 173:3, 1654-1667, doi: 10.1007/s11232-
012-0139-x, arxiv.org/abs/1406.5698.

\bibitem[Magazev et al. (2008)]{39}
A. A. Magazev, I. V. Shirokov, Yu. A. Yurevich, Integrable magnetic geodesic flows on Lie groups, {\bf 2008}, TMF, 156:2, 189-206; {\em Theoret. and Math. Phys}. {\bf 2008} 156:2, 1127-1141. doi.org/10.4213/tmf6240

\bibitem[Magazev (2021)]{40}
A.A.Magazev, Constructing a Complete Integral of the Hamilton-Jacobi Equation on
Pseudo-Riemannian Spaces with Simply Transitive Groups of Motions. {\em Math. Phys. Anal
Geom.} 24, 11, {\bf 2021}. https://doi.org/10.1007/s11040-021-09385-3

\bibitem[Obukhov (2023)]{41}
Obukhov V. V.. Hamilton-Jacobi and Klein-Gordon-Fock equations for a charged test
particle in space-time with simply transitive four-parameter groups of motions. J. Math.
Phys.{\bf 2023} 64, 093507; doi: 10.1063/5.0158054

\bibitem[Obukhov (2022)]{42}
Obukhov V.V. Algebras of integrals of motion for the Hamilton-Jacobi and Klein-
Gordon-Fock equations in spacetime with a four-parameter groups of motions in
the presence of an external electromagnetic field. {\em J. Math. Phys.} {\bf 2022}, 63, Issue 2.
https://doi.org/10.1063/5.0080703

\bibitem[Obukhov (2021)]{43}
 Obukhov V.V. Algebra of symmetry operators for Klein-Gordon-Fock Equation. {\em Symmetry.}
{\bf 2021}, 13, 727 (15p.). https://doi.org/10.3390/sym13040727.

\bibitem[Obukhov (2022)]{44}
Obukhov V.V. Algebra of the symmetry operators of the Klein-Gordon-Fock equation for the case when groups of motions G3 act transitively on null subsurfaces of spacetime.
{\em Symmetry}. {\bf 2022}, 14, (346). https://doi.org/10.3390/sym14020346

\bibitem[Komrakov (2001)]{45}
 B. B. Komrakov, Einstein-Maxwell equation on four-dimensional homogeneous spaces, Lobachevskii {\em J. Math.} {\bf 2001} 8, 33-165.

\bibitem[Calvaruso et al. (2015)]{46}
Calvaruso G., and Fino A. {\bf 2015}. Four-dimensional pseudo-Riemannian homogeneous Ricci solitons. International {\em Journal of Geometric Methods in Modern Physics}, 12(05), 1550056.
doi.org/10.48550/arXiv.1111.6384

\bibitem[Calvaruso et al. (2014)]{47}
Calvaruso, Giovanni, and Amirhesam Zaeim. A complete classification of Ricci and Yamabe solitons of non-reductive homogeneous 4-spaces. {\em Journal of Geometry and Physics} 80 (2014): 15-25. doi.org/ 10.1016/j.geomphys.2014.02.007

\bibitem[Ugur Camci (2024)]{48}
Ugur Camci. Noether Symmetry Analysis of the Klein-Gordon and Wave Equations in Bianchi I Spacetime. {\em Symmetry}, {\bf 2024}, 16(1), 115; doi.org/10.3390/sym16010115

\bibitem[Ghezelbash et al. (2022)]{49}
Ghezelbash A. M.Bianchi IX geometry and the Einstein-Maxwell theory {\em Class. Quantum Grav.} 39 {\bf 2022} 075012 (36pp) doi.org/10.1088/1361-6382/ac504e

\bibitem[Obukhov (2022)]{50}
V.V.Obukhov. Maxwell Equations in Homogeneous Spaces for Admissible Electromagnetic Fields". Universe. {\em 8}, (245), (2022). https://doi.org/10.3390/universe8040245

\bibitem[Obukhov (2022)]{51}
V.V.Obukhov. Maxwell Equations in Homogeneous Spaces with Solvable Groups of Motions. {\em Symmetry}. {\em 14}, 2595, (2022).doi.org/10.3390/sym14122595

\bibitem[Obukhov (2023)]{52}
 V. V. Obukhov. "Exact Solutions of Maxwell Equations in Homogeneous Spaces with the Group of Motions G3(IX). Axioms {\em 12}, {\em 135}, (2023). doi.org/10.3390/axioms12020135

\bibitem[Obukhov (2023)]{53}
V.V.Obukhov, Exact Solutions of Maxwell Equations in
Homogeneous Spaces with the
Group of Motions G3(VIII). {\em Symmetry.} {\em 15}, 648, (2023). doi.org/10.3390/sym15030648

\bibitem[Obukhov et al. (2024)]{54}
Obukhov V. V. , Chervon S. V., and Kartashov D. V.
International Journal of Geometric Methods in Modern PhysicsOnline ReadyNo Access
Solutions of Maxwell equations for admissible electromagnetic fields, in spaces with simply transitive four-parameter groups of motions. {\em IJGMMP}, {\bf 2024}. doi.org/10.1142/S0219887824500920

\bibitem[Kasner (1921)]{55}
Kasner, Edward. Geometrical Theorems on Einstein Cosmological Equations.{\em American Journal of Mathematics} {\bf 1921}, 43, no. 4: 217-21. doi.org/10.2307/2370192.

\bibitem[Mitter (1931)]{56}
Mitter O.K. On a solution of einstein's gravitational equations $G_{\mu\nu} =0$ symmetrical an axis. {\em Tohoku Math. Journ}. {\bf 1931}, 34.

\bibitem[Petrov(2019)]{57}
Petrov A. Z. New methods in General Relativity (in Russian,),{\em  M. KRASAND,} {\bf 2019} 496 p. ISBN 978-5-396-00884-7

\bibitem[Taub (1951)]{58}
Taub, A. H. Empty Space-Times Admitting a Three Parameter Group of Motions. {\em  Annals of Mathematics}. {\bf 1951}, 53, no. 3: 472-90. doi.org/10.2307/1969567.

\bibitem[Rosen (1962)]{59}
Rosen G. Symmetries of the einstein-maxwell equations, {\em J. Math. Phys.} {\bf 1962}, 3, 2.

\bibitem[Landau et al. (1988)]{60}
Landau, L.D.; Lifshits, E.M. Theoretical Physics, Field Theory, 7th ed.; {\em Science, C.,
Ed.; Nauka: Moskow, Russia,} 1988; Volume II, 512p. ISBN 5-02-014420-7.

\bibitem[Bagrov et al. (1983)]{61}
Bagrov, V. G.Obukhov, V. V.Classes of Exact Solutions of the Einstein-Maxwel-Equations {\rm Annalen der Physik.} {\bf 1983}. {\bf B. 495},  H. 4–5, (181–188).doi:10.1002/andp.19834950402.

\bibitem[Bagrov et al. (1986)]{62}
Bagrov, V. G.Obukhov, V. V., Shapovalov, A. V. Special Stackel Electrovac Spacetimes {\rm Pramana, Journ. of Phys}. {\bf 1986}, {\bf 26}, 2, (93–108). doi: 10.1007/BF02847629.


\bibitem{1111}
Bagrov, V. G.Obukhov, V. V., Shapovalov, A. V. Special Stackel Electrovac Spacetimes {\rm Pramana, Journ. of Phys}. {\bf 1986}, {\bf 26}, 2, (93–108). doi: 10.1007/BF02847629.

\end{thebibliography}






\end{document}